Vibrations of Periodically Poled Lithium Niobate Bar with 0.3-mm Long Domains.

Igor Ostrovskii [a], Andriy Nadtochiy [a,b], Lucien Cremaldi [a]

[a] University of Mississippi, Department of Physics and Astronomy, Oxford, MS 38677, U.S.A.

[b] Taras Shevchenko National University of Kyiv, Kyiv 01601, Ukraine.

**Abstract**

The three dimensional vibrations in a periodically poled ZX-cut Lithium Niobate thin bar with 0.3-mm long domains are considered. The acoustical vibrations may be excited by the 1) longitudinal acousto-electric current when a radio frequency voltage is parallel to the x-axis, and 2) transverse electric field when a radio frequency voltage is parallel to the z-axis. The computations by the Final Element Method reveal all the three displacements along the x, y, and z crystallographic axes. The amplitudes may be different for two types of vibration excitations. The positions of peaks in admittance versus frequency correlate with the frequencies of maximum acoustic amplitudes. The superlattice with 0.3-mm long domains along the x-axis is fabricated in the z-cut 0.5-mm-thick wafer. The experimental data on the radio frequency admittance versus frequency is in agreement with the corresponding theoretical computations.

**Introduction**

The acoustic properties of periodically poled phononic superlattices (PS) are extensively investigated due to their attractive physical characteristics and various potential applications. The ferroelectric based superlattices including those fabricated in the periodically poled lithium niobate and lithium tantalite are widely used for applications in telecommunications, Ultrasonics [1-5] and acousto-optics for laser technology [4]. The megahertz and higher frequency ranges are often used in telecommunications including cell phones [2]. That is why the properties of multidomain PS are important for those frequencies. The bulk type PS usually has its piezoelectric coefficient as a periodically poled along one axis while host crystal itself may be considered as infinite along other two axes. Next step to miniaturization is a wafer type PS. For instance, periodically poled ferroelectric plate [6 - 8]. In all types of PS mentioned above, it is usually considered one dominated acoustic displacement along certain crystallographic axis. However, in the real applications such as ultrasonic transducers [9], electromechanical actuators, radio-frequency filters [2], acousto-optic devices [4] and others no one dimension is of infinite order. Thus for theoretical and experimental researches, it is highly recommended to take into account all three dimensions for making researches closer to the real applications. One of the reasons is a respectively high Poisson's ratio of solids that is from about 0.2 to 0.4 for different materials. Lithium niobate has Poisson's ratio of 0.25 and that implies near 25% displacement in the normal directions with respect to an original axis along which a 100% displacement is introduced. The important consequence of such vibrations is actually tree-dimensional character of the acoustic displacements in a PS. The three dimensional vibrations with all three amplitude components along the x, y, and z axes in turn may be a basis for new solid state devices. For instance, one can mention a new generation of ultrasonic transducers with the variable direction and polarization of the mechanical displacements.

In this work, we demonstrate an existence of three-dimensional vibrations along the x, y, z axes in a periodically poled system consisting of 0.3-mm-long ferroelectric domains in a lithium niobate thin bar.

**Computational and experimental results**.

The multidomain periodically poled structure under consideration is presented in Fig.1.

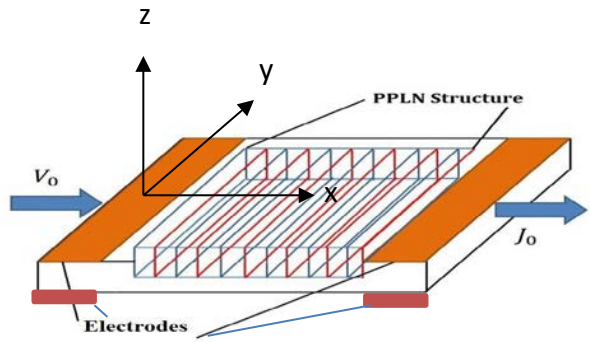

Fig. 1. Periodically poled phononic superlattice in a thin lithium niobate wafer.
The periodically poled ferroelectric domains are shown by the bars between red and blue planes. The neighboring domains are polarized along the (+z) axis direction or in opposite (–z) direction. The metal electrodes are for exciting vibrations. The dimension along the y-axis is not infinite.

The radio frequency voltage *Vo* is applied to top metal electrodes for exciting acoustic vibrations in the PS, and the radio frequency current *Jo* may be theoretically calculated along with the acoustic displacements, and further experimentally measured as a function of frequency.

The problem of finding the vibration amplitudes is solved by using the equations of motion and electrodynamics for piezoelectric media, equations (1) to (4) along with the harmonic solutions for the displacements ($u$) as in equation (5):

$$\rho \frac{\partial^2 u}{\partial t} = \nabla \cdot T + f_V \quad (1)$$

$$T = c^E S - e^t E \quad (2)$$

$$D = eS + \varepsilon^S E \quad (3)$$

$$\nabla \cdot D = 0 \quad (4)$$

$$u = u(x, y, z) \cdot e^{j\omega t} \quad (5)$$

In the equations, $T$ is stress tensor, $\rho$ is crystal density, $f_V$ is external force acting on crystal, $c^E$ is elastic module tensor, $S$ is strain tensor, $e^t$ stands for piezoelectric constant tensor, $E$ is electric field, $\varepsilon^s$ is dielectric constant tensor, $D$ is electrical induction. Further, the problem is computed by using the Finite Element Methods (FEM). The details with regard to FEM-computation for a piezoelectric wafer are described in the reference [6]. The connection between the $J_o$ current through PS and the amplitude $A_o$ of an acoustic vibration is given by equation (6) from reference [9].

$$A_0 = J_0 \frac{e(1+K^2)(1+Z_0)^{-1}}{2kc\varepsilon\omega(k_0^2/k^2)\sin(kNd/\sqrt{1+K^2})} \quad (6)$$

Where, $K^2$ is squared effective piezoelectric coefficient, $k$ is wave-number of an acoustic mode in PS and $k_0$ is wave-number of the same acoustic mode in a crystal without PS, $N$ denotes the number of inversely poled domains in a PS, $d$ stands for a length of single domain along the x- axis, and $Z_0$ is ratio of the acoustic impedance of a load over those of a crystal with PS. If a PS is not acoustically connected to any media, then $Z_0 = 0$.

The FEM computations are made for the periodic structure with the following parameters: domain length d = 0.3 mm along the x-axis with total number of domains N = 70, crystal dimensions are 23mm (along the x-axis) x 14 mm (along the y-axis) x 0.5 (along the z-axis). Thus, the multidomain area is 21x14mm$^2$, and the metal electrodes are 1 mm along the x-axis and 14 mm along the y-axis. The FEM mesh is 350x10x5 elements, and the Finite Element type is a quadratic hexahedral 20-node element. The uniform domains of both polarizations have their length of 0.3 mm along the x-axis. The computed amplitudes of the mechanical displacements are presented in Fig. 2 for excitation by the current *Jo*. The voltage *Vo* is applied between the left and right electrodes in Fig. 1 and so is parallel to the x-axis.

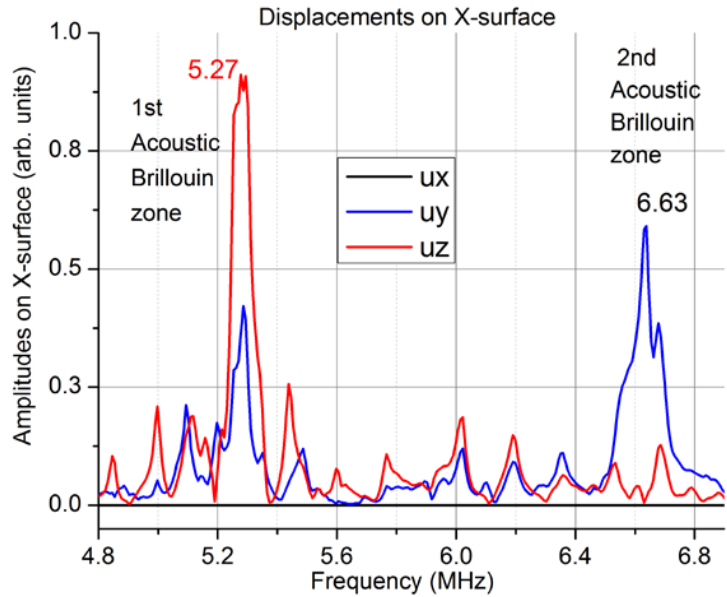

Fig. 2. The FEM computation for the acoustic amplitudes on the x-surface under excitation by an acousto-electric current *Jo*. Voltage *Vo* is along the x-axis.

The excitation of vibrations can also be done by applying an rf-voltage *Vo* to the two electrodes located on the upper and lower surfaces of the wafer, then *Vo* is parallel to the z-axis. This configuration is usual for an application of so-called delay line. The input pulse/voltage is applied to the top and bottom electrodes on the left side in Fig. 1 and an output pulse/signal is detected by the two electrodes on the right side of crystal in Fig.1. The results are presented in Fig. 3.

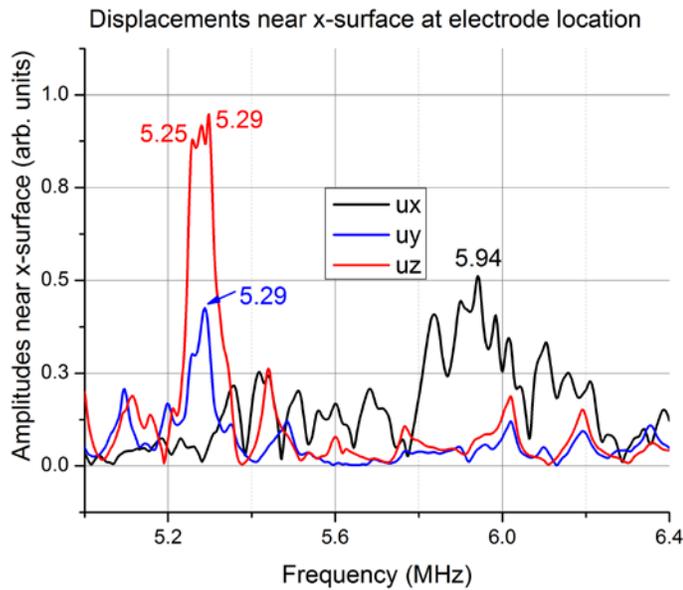

Fig. 3. The FEM computation for the acoustic amplitudes near the x-surface under excitation by a radio frequency voltage applied along the z-axis.

The results of figures 2 and 3 show that one can excite all the three vibrational components along x, y, or z axes. Actually by applying a proper excitation and frequency, it is possible to get the vibrations of dominated amplitudes, that is *Ux* or *Uy*, or *Uz*. For instance, the excitation by the current *Jo* at 6.63 MHz gives mainly *Uy* vibration. However, the excitation by the voltage *Vo* parallel to the z-axis at 5.94 MHz gives mainly *Ux* vibration. Overall result is as follow. The maximal amplitudes are within the frequency range of 5.2 to 5.4 MHz, which corresponds to the first acoustic Brillouin zone. The second frequency range of maximal amplitudes is 5.94 to 6.63 MHz. These conclusions may be independently verified by computing and experimental measurements of a radio frequency admittance Y(F). The admittance is proportional to the current *Jo*, which in turn is connected to the vibration amplitude *Ao* as in the equation (6). The results of FEM computation of Y(F) are presented in Fig. 4.

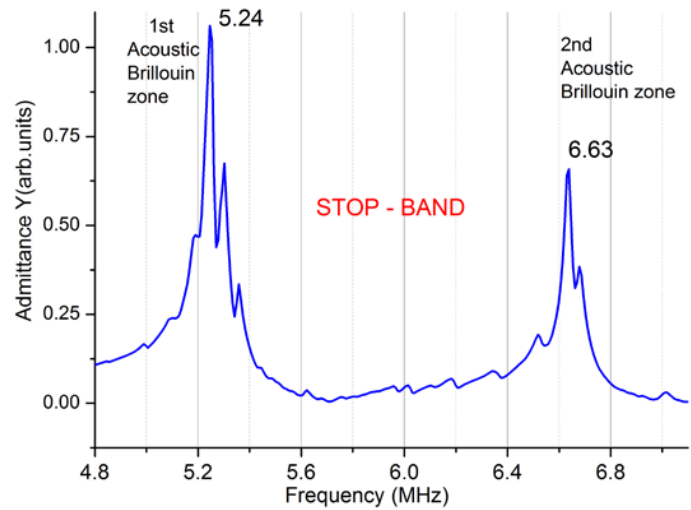

Fig. 4. The FEM computation for rf admittance Y(F) under excitation

by current *Jo*, voltage *Vo* is parallel to the x-axis.

The experimental measurements are taken with the sample ZX-LN-MD1B that has 70 periodically poled domains. The acousto-electric quality factor is Q = 250. The micro-picture of the multidomain structure taken through the polarized optical microscope is shown in Fig. 5.

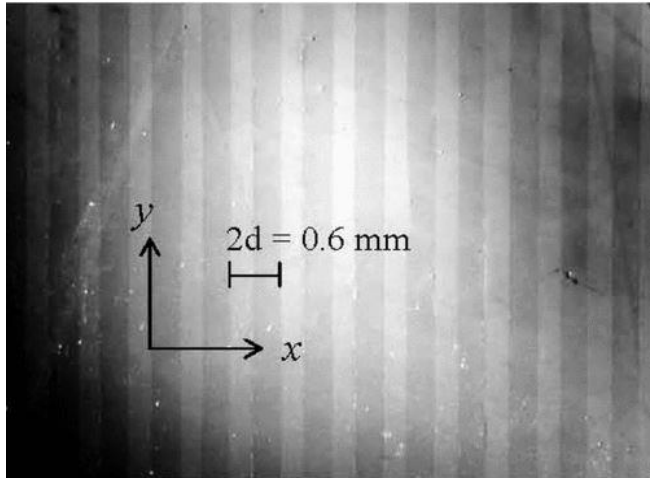

Fig. 5. Micro-picture of the multidomain structure with 0.3-mm long domains, ZX-LN-MD1B.

In Fig. 6, we present the results of experimental measurements of the PS-admittance versus frequency from the sample ZX-LN-MD1B at room temperature.

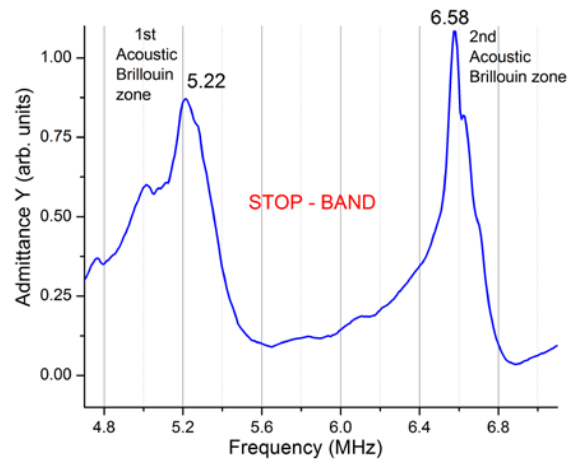

Fig. 6. Experimentally measured rf admittance Y(F) from ZX-LN-MD1B sample under excitation by the current *Jo*, voltage *Vo* is parallel to the x-axis.

.

**In Conclusion**

The finite element method computations yield the acoustic displacements and acousto-electric admittance versus frequency. There are two methods to exciting vibrations: 1) excitation by the acousto-electric current *Jo* when rf voltage *Vo* is parallel to the x-axis, and 2) excitation by a rf voltage normal to a wafer surface when *Vo* is parallel to the z-axis. The acoustic displacements have three components along the x, y and z axes under both types of vibration excitation. The vibration amplitudes are minimized within the stop-band frequencies, which may vary for different displacement components and different type of excitation. More pronounced stop band appears when the vibrations are excited by the longitudinal current *Jo* that flows across the length of all multidomain structure along the x-axis. The comparison of the figures 2 and 3 along with the figures 4 and 6 allow conclude that all three polarizations of the acoustical vibrations may be excited in a periodically poled multidomain superlattice in a thin lithium niobate wafer. The individual displacements along the x, y, or z axes may be generated by a proper choice of frequency and method of vibration excitation. The acousto-electric resonances in the admittance Y(F) correspond to the maxima in acoustic amplitudes. The results obtained may be used for development of new applications such as multi-displacement ultrasonic transducers and actuators, acousto-electric filters and oscillators.